\documentclass[12pt]{iopart}

\usepackage{iopams}
\usepackage{epsfig}
\begin{document}
\def\be {\begin{equation}}
\def\ee {\end{equation}}
\def\nn {\nonumber}
\def\bea {\begin{eqnarray}}
\def\eea {\end{eqnarray}}
\def\k {|{\vec k}|}
\def\q {|{\vec q}|}
\def\ks {{|{\vec k}|}^\ast}
\def\qs {|{{\vec q}|}^\ast}
\def\mqs {M_q^\ast}
\def\eqs {E_q^\ast}
\def\eq {E_q}
\def\mks {M_k^\ast}
\def\ek {E_k}
\def\eks {E_k^\ast}
\def\mp {m_\pi}
\newcommand{\bef}{\begin{figure}}
\newcommand{\eef}{\end{figure}}
\newcommand{\ra}{\rightarrow}
\newcommand{\gm}{\gamma^\mu}
\newcommand{\gf}{\gamma^5}
\newcommand{\N}{\bar{N}}
\newcommand{\del}{\partial}
\def\bt{{\bf{\tau}}}
\def\ba{{\bf{a_1}}}
\def\br{{\bf{\rho}}}
\def\bp{{\bf{\pi}}}
\def\cl{{\cal{L}}}
\def\prl{{\em Phys. Reports }}
\def\prl{{\em Phys. Rev. Lett. }}
\def\prc{{\em Phys. Rev. {\bf C} }}
\def\prd{{\em Phys. Rev. {\bf D} }}
\def\jap{{\em J. Appl. Phys. }}
\def\ajp{{\em Am. J. Phys. }}
\def\nima{{\em Nucl. Instr. and Meth. Phys. {\bf A} }}
\def\npa{{\em Nucl. Phys. {\bf A}}}
\def\npb{{\em Nucl. Phys. {\bf B}}}
\def\epjc{{\em Eur. Phys. J. {\bf C}}}
\def\epja{{\em Eur. Phys. J. {\bf A}}}
\def\plb{{\em Phys. Lett. {\bf B}}}
\def\mpla{{\em Mod. Phys. Lett. {\bf A}}}
\def\pr{{\em Phys. Rep.}}
\def\zpc{{\em Z. Phys. {\bf C}}}
\def\zpa{{\em Z. Phys. {\bf A}}}
\def\app{{\em Acta Physica Polonica {\bf B}}}
\def\rnc{{\em Riv. Nuovo Cimento}}
\newcommand{\eovern}{\left<E\right>/\left<N\right>}
\title{The QCD Equation of State with Thermal Properties of $\phi$ mesons}
\author{Raghunath Sahoo$^{1}$, 
Tapan K. Nayak$^2$, Jan-e Alam$^2$, Sonia Kabana$^1$\footnote{Speaker}, Basanta K. Nandi$^3$ 
and D P Mahapatra$^4$}

\address{$^1$SUBATECH, 4, Rue Alfred Kastler, BP 20722 - 44307 Nantes Cedex 3, France}
\address{$^2$Variable Energy Cyclotron Centre, 1/AF Bidhan Nagar, Kolkata 700064, India}
\address{$^3$Indian Institute of Technology Bombay, Powai, Mumbai 400076, India}
\address{$^4$Institute of Physics, Sachivalaya Marg, Bhubaneswar 751005, India}

\begin{abstract}
In this work a first attempt is made 
to extract the Equation of State (EoS) using experimental results of the $\phi$ meson 
produced in nuclear
collisions at AGS, SPS and RHIC energies.
The data are confronted to simple thermodynamic expectations and lattice results.
The experimental data indicate a first order phase transition, with a mixed phase stretching
energy density between $\sim$ 1 and 3.2 GeV/$fm^3$.
\end{abstract}

\section{Introduction}
Quantum chromodynamics (QCD) predicts a phase transition from hadronic matter to the
quark and gluon state at a critical temperature $T\sim 200$ MeV \cite{lQCD}.
There are experimental evidences that nuclear collisions 
at ultrarelativistic energies induce this QCD phase transition
and create a new state of matter where the properties of 
the matter are governed by quarks and gluons - such a system is called Quark Gluon Plasma (QGP).
Current lattice QCD calculations indicate that the 
order of this transition  depends on the quark masses,
as well as on the baryochemical potential ($\mu_B$).
In particular, a first order phase transition is predicted for large baryochemical potentials,
while at zero and small baryochemical potentials a "cross over" is predicted.
When studying nuclear collisions at ultra-relativistic energies
the baryochemical potential at midrapidity decreases with increasing energy.
Therefore,  lattice QCD indicates that the order of the transition changes with 
energy ($\mu_B$) of the collision.  However, lattice calculations for non-zero $\mu_B$  
have large technical uncertainties and the best estimate of lattice remains at $\mu_B$=0.

\noindent
~~~In a first order phase transition the pressure increases with increasing temperature, until 
the transition temperature $T_c$ is reached, 
then remain constant during the mixed phase, and continue to increase after the end of the 
mixed phase. 
Related to this picture  L. Van Hove \cite{vanHove} suggested 
to identify the deconfinement transition
in high energy proton-antiproton collisions, looking at the variation 
 of average transverse momentum ($\left<p_T\right>$) of hadrons
 as a function of the
hadron multiplicity at midrapidity ($dN/dy$) and
 searching for an increase of $\left<p_T\right>$
 followed by a  plateau-like behavior and again a subsequent increase.
The $\left<p_T\right>$ 
 is expected to reflect  the thermal freeze-out temperature $(T_f)$ of hadrons 
and a flow component which can be related  to the initial pressure,
similarly the hadronic multiplicity reflect the entropy density of
the system.
While the purely thermal component of $\left<p_T\right>$
 can not be related to the initial temperature which remains unmeasurable above $T_c$,
however the flow component in the inverse slope can
reflect the plateau of the pressure during mixed phase.

\noindent
~~~It has been observed that the $\left<m_T\right>$ of pions, kaons and protons
as a function of $dN_{ch}/dy$
shows a Van-Hove-like behaviour as explained above
for a wide range of collision energy and for the same centrality 
\cite{bm,marek}. 
Hydrodynamic calculations assuming a first order transition could reproduce
these data \cite{bm,kodama}. In this work, we study for the first time
the variation of the inverse slope, $T_{\textrm{eff}}$ extracted from the
$p_T$ distribution of the $\phi$ -meson as a function of the
initial energy density, $\epsilon_{Bj}$  evaluated within the framework of 
Bjorken's hydrodynamical model~\cite{bjorken} and  $\sqrt{s_{NN}}$
for energies spanning from AGS, SPS to RHIC. 
The $\phi$-meson is of special interest \cite{bedanga_sqm2008}
because of its small hadronic rescattering cross section of $\sigma$($\phi$N)=10 mb
causing it to decouple earlier than other hadrons.
Furthermore, due to its life time of $\sim$ 45 fm/c, its main decay product $(K^+K^-)$
suffer  less  rescattering. 
Experimental results from Au+Au collisions at RHIC energies indicate  that the $\phi$ 
has a higher thermal freeze out temperature as compared to pions, kaons and protons.
In particular their thermal freeze out temperature is within errors compatible with the
chemical freeze out temperature of hadrons and the critical temperature.
Another 
important experimental fact is the observation of a scaling of the elliptic
flow of  hadrons
 as a function of the transverse momentum when divided with the number of valence quarks.
This observation is interpreted as an indication that elliptic flow builds in the partonic 
phase and can therefore be reflecting the initial
conditions e.g. the initial pressure.
Therefore the $\phi$  and its flow phenomena
 are particularly interesting probes for studying the EoS
and the nature of the phase transition.

\noindent
~~~Furthermore, we study here for the first time the inverse slope as a function of the 
initial energy density, $\epsilon_{Bj}$.  This is an important new
feature of such studies, because it connects the inverse slope with
a parameter characterizing the initial state of the collision build up after$\sim$ 1 fm/c
and which reflects at the same time the collision energy, the stopping and the impact parameter 
of the collision.  For example, at a given
 $\sqrt{s_{NN}}$, different energy densities could be achieved
by changing the colliding nuclei species or the impact parameter. 
The
 $\epsilon_{Bj}$  is a meaningfull parameter
 also if equilibrium is not reached. It can be 
directly compared to the critical energy density $\epsilon_c$ 
obtained in lattice QCD calculations- $\epsilon_c\sim 1$ GeV/fm$^3$.
In the following the $\phi$ data will be analysed and confronted to
simple thermodynamic expectations which relate to the Van Hove signature and
to lattice QCD predictions. 

\section{$\phi$ as a probe of the order of the phase transition}

\noindent
To study the dependence of the inverse slope of the $\phi$ meson 
on the collision energy and $\epsilon_{Bj}$, we have compiled the
inverse slope of the transverse mass spectra ($m_T=\sqrt{ m_\phi^2 + p_T^2}$)
called "effective temperature" $T_{\textrm{eff}}^{\phi}$ using 
data at mid-rapidity in the low $p_T$ domain from 
AGS \cite{agsPhi}-SPS \cite{floris,na38,na50,na49,spsPhi} to RHIC \cite{rhicPhi}.
The
 inverse slope has been estimated  using an exponential function of the type:
$\frac{1}{m_T}\frac{dN}{dm_T} = A.~ exp~(\frac{m_T}{T_{eff}})$\label{exp}. The
 use of slightly different fit functions
leads
 to differences of the order of 10 MeV, which are small as compared to the
experimental errors on the inverse slopes \cite{na38}.
The
 inverse slope of the $\phi$ meson extracted through the above fit includes
a thermal component and a non-thermal component due to the collective transverse flow \cite{nu}.
At
 low $p_T$ (in non-relativistic domain, $p_T \ll m$),
$
T_{eff} = T_{th} + \frac{1}{2}m\left<\beta_T\right>^2,
$
\label{teff}
where
 $\beta_T$ is the collective transverse velocity. 
We have used the above equation to 
estimate 
$T_{\textrm{th}}^{\phi}$ from $T_{\textrm{eff}}^{\phi}$, as a first approximation.
The values of $\left<\beta_T\right>$ as a function of $\sqrt{s_{NN}}$ has been obtained 
from Ref.\cite{nu} and as a function of centrality for $\sqrt{s_{NN}}=200$ GeV AuAu 
collisions from Ref.\cite{starPRL}.

\begin{figure}
\begin{center}
\resizebox{1.0\columnwidth}{!}{%
\includegraphics{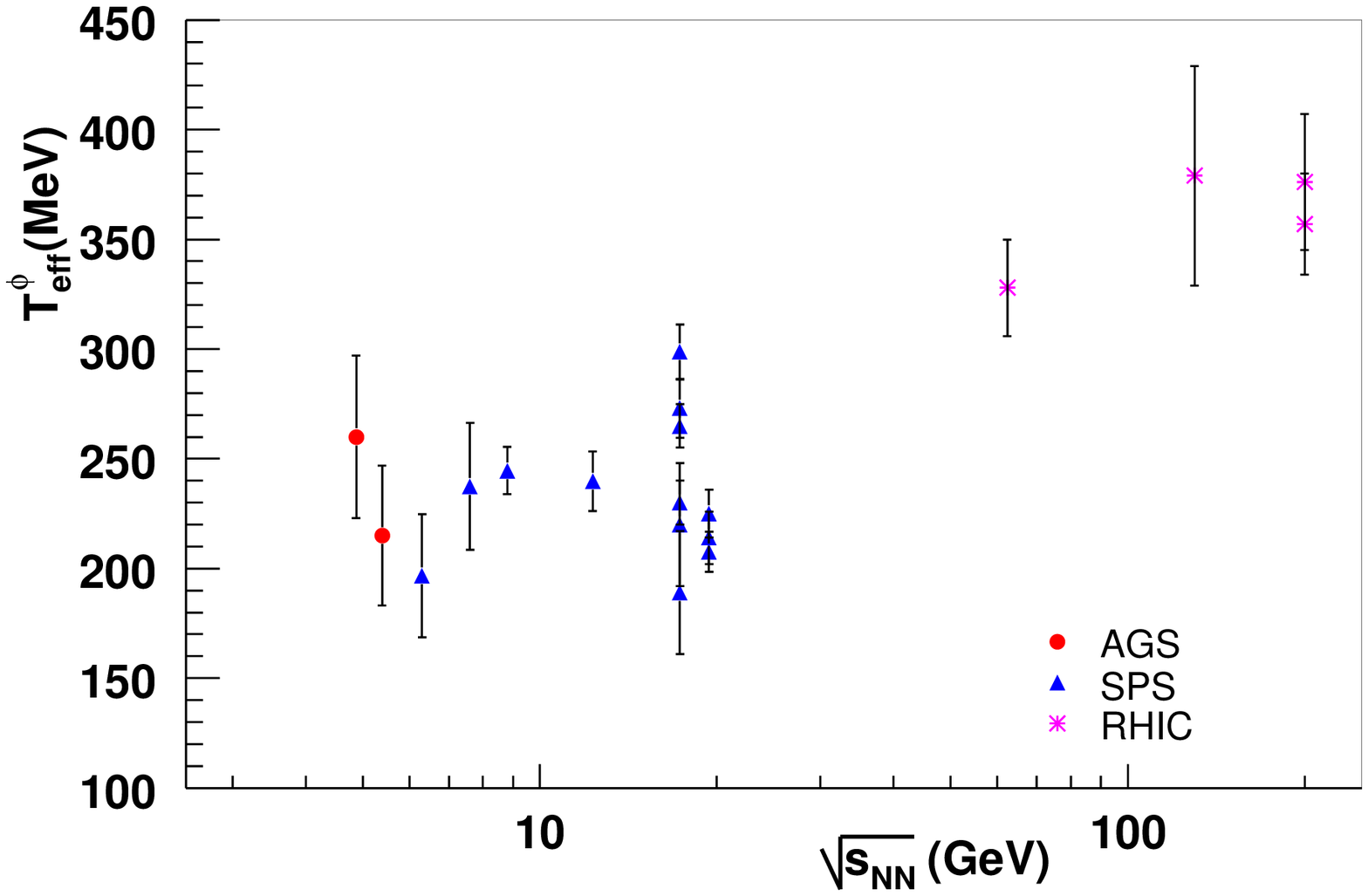}
\includegraphics{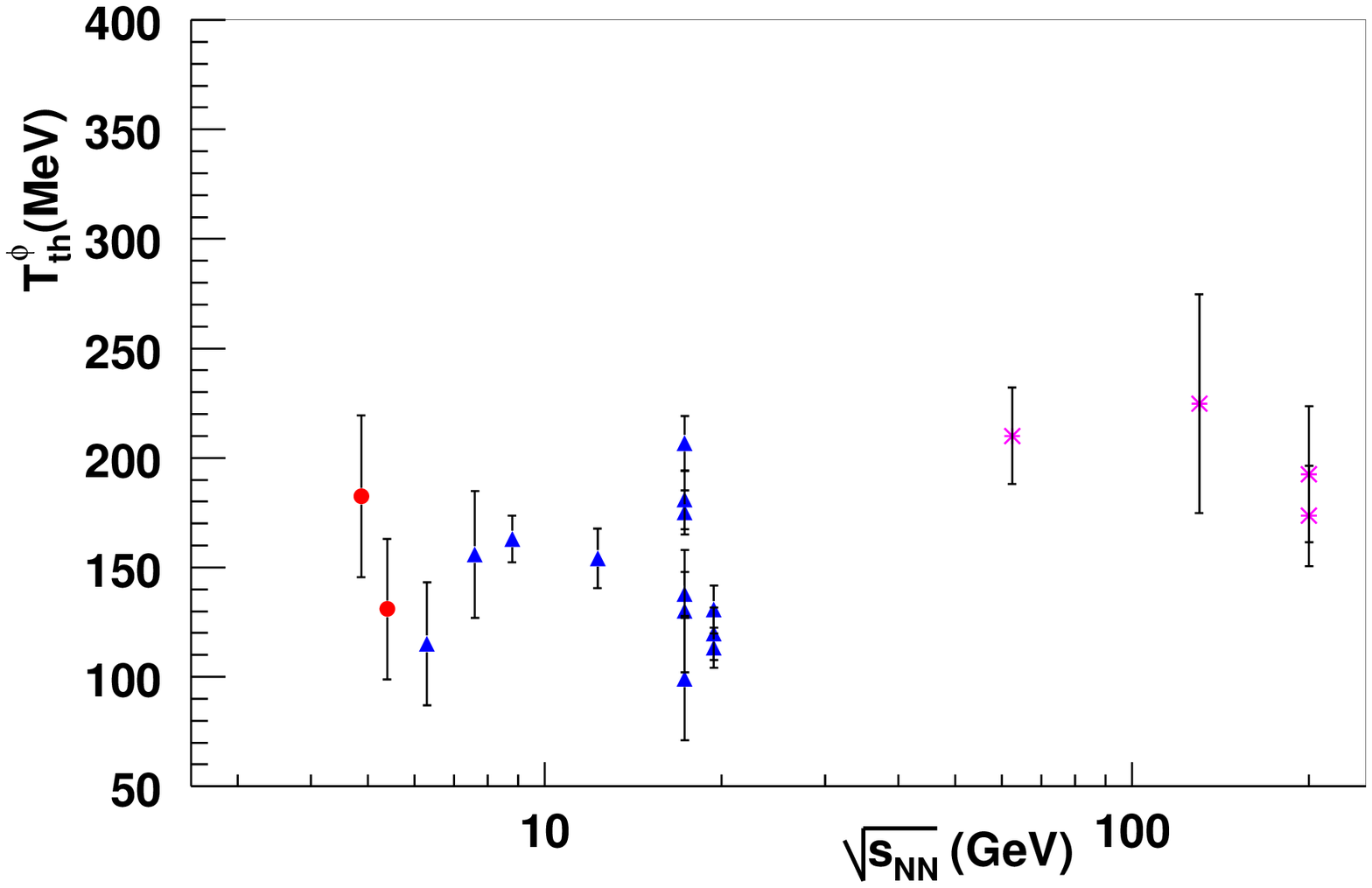}}
\caption{$T_{\textrm{eff}}^{\phi}$ (left ) and $T_{\textrm{th}}^{\phi}$ (right) as 
a function of $\sqrt{s_{NN}}$ from AGS-SPS to RHIC.}
\label{withCoM}
\end{center}
\end{figure}

\noindent
~~~In figure \ref{withCoM} the effective temperature 
of $\phi$, $T_{\textrm{eff}}^{\phi}$ (left) and the thermal component of the inverse slope 
  $T_{\textrm{th}}^{\phi}$ (right)
 are shown as a function of the collision energy $\sqrt{s_{NN}}$.
It is observed that from AGS to SPS energies these
observables remain almost unchanged, showing a plateau-like structure. 
Going from 
SPS to RHIC energies, the inverse slope $T_{\textrm{eff}}^{\phi}$ exhibits a sudden jump
while an increase is still observed in the  $T_{\textrm{th}}^{\phi}$ 
component.
This may be due to an imperfect transverse flow component subtraction at RHIC
or other effect.
It 
is observed that the $T_{\textrm{th}}^{\phi}$
is reaching at RHIC  values compatible within errors with $T_c$.
The
 observed
 plateau of $T_{\textrm{eff}}^{\phi}$ is a signature of a coexisting phase of quarks, gluons and hadrons
for a first order phase transition,
during which the initial pressure remains constant.
The subsequent
increase of $T_{\textrm{eff}}^{\phi}$ with $\sqrt{s_{NN}}$ at top RHIC energies indicates 
the end of the mixed phase and the entering into a pure QGP phase.

\begin{figure}
\begin{center}
\resizebox{0.8\columnwidth}{!}{%
\includegraphics{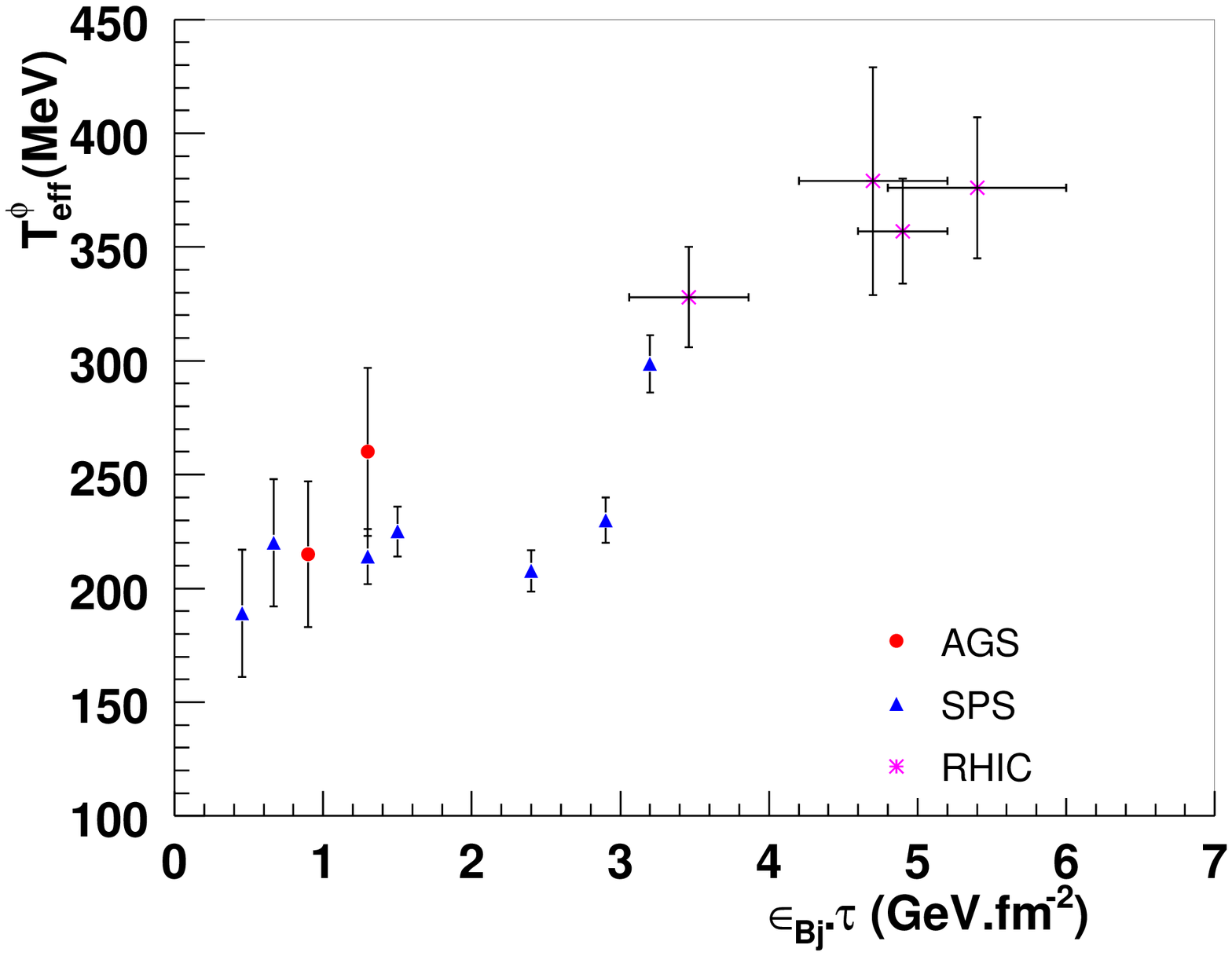}
\includegraphics{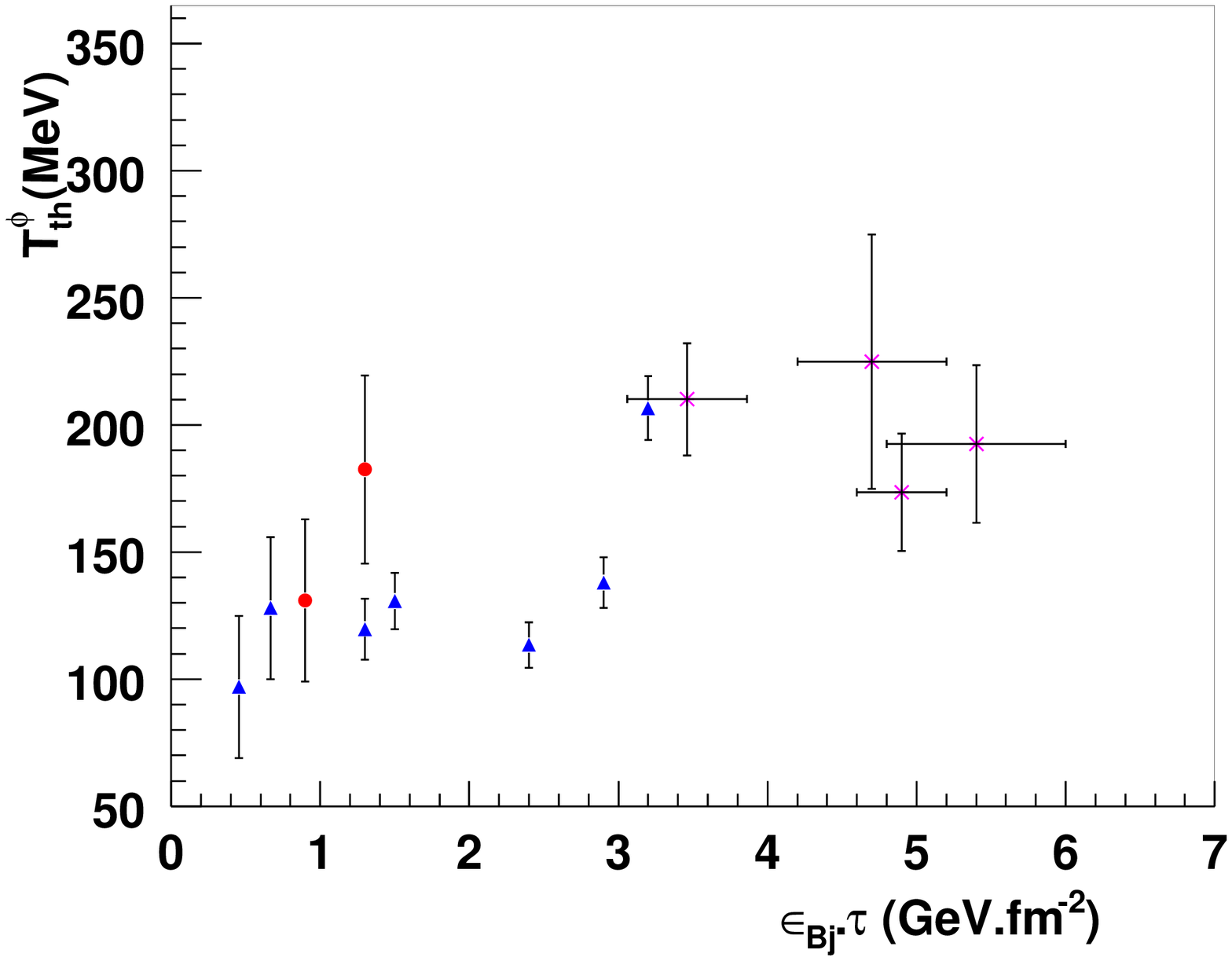}}
\caption{$T_{\textrm{eff}}^{\phi}$ and $T_{\textrm{th}}^{\phi}$ as 
a function of $\epsilon_{\textrm{Bj}}.\tau$ from AGS-SPS to RHIC.}
\label{withEDensity}
\end{center}
\end{figure}

\noindent
~~~Now we study the dependence of the inverse slope and its thermal component
on  the initial energy density, $\epsilon_{Bj}$, estimated as
$ \epsilon_{Bj} = \left<\frac{dE_T}{dy}\right>\frac{1}{\tau\pi R^2} = \left<\frac{dN}{dy}\right>\left<m_T\right> \frac{1}{\tau\pi R^2}
$
where $R = R_0 A^{1/3}$ and $A\sim \frac{N_{part}}{2}$.
From
 the experimental measurements of $dE_T/dy$ and $dN/dy$ with $\left<m_T\right>$,
the observable $\epsilon_{Bj}$ could be estimated for all centralities 
and center of mass energies at mid-rapidity.

\noindent
~~~In figure \ref{withEDensity} 
 $T_{\textrm{eff}}^{\phi}$ (left) and $T_{\textrm{th}}^{\phi}$ (right)
 are shown as a function of $\epsilon_{Bj}.\tau$. Note that the formation time, 
$\tau$ is model-dependent and in general, in subsequent discussions we also assume
$\tau \sim 1$ fm/c.
The above result reflects the properties of equation of state.
We observe a plateau in $T_{\textrm{eff}}^{\phi}$ stretching between $\epsilon_{Bj}$
 $\sim$ 
1 and 3.2 GeV/$fm^3$, and increasing suddently above 3.2 GeV/$fm^3$.
This 
behaviour as already discussed, suggests a first order transition,
however from figure 2 we can now infer that the
 mixed phase is stretching between 1 and 3.2 GeV/$fm^3$.
The
use of the $\epsilon_{Bj}$ scale, allows us to establish
here for the first time the $\epsilon_{Bj}$ range of the mixed phase.
More
 data on the $\phi$ at  $\epsilon_{Bj}$ below 1 GeV/$fm^3$ are needed
to establish the increase of $T_{\textrm{eff}}^{\phi}$ up to 1 GeV/$fm^3$,
 as seen in other hadrons. 
The
 increase of the inverse slope at RHIC energies again indicate a pure phase of QGP, as 
is expected from a first-order phase transition. 
The thermal component $T_{\textrm{th}}^{\phi}$ shows also a plateau while a smaller increase
is still observed at 3.2 GeV/$fm^3$.

\noindent
~~~ The flow component of the inverse slope of
the $\phi$ reflecting the initial pressure, it is of interest to look directly observables
linked to this initial pressure like the  transverse flow velocity,
$\beta_T$ for all hadrons and the elliptic flow $v_2$ as
a function of collision energy.
The transverse flow  velocity
$\beta_T$  shows exactly the same characteristic behaviour as the
$T_{\textrm{eff}}^{\phi}$ as a function of collision energy, namely increase,
a plateau and subsequent increase at RHIC \cite{nu}.
A similar pattern is suggested for the elliptic flow $v_2$ as a function of collision energy
\cite{lokesh_sqm2008}.

\section{Comparison of $\phi$ data with lattice predictions}

\begin{figure}
\begin{center}
\resizebox{1.1\columnwidth}{!}{%
\includegraphics{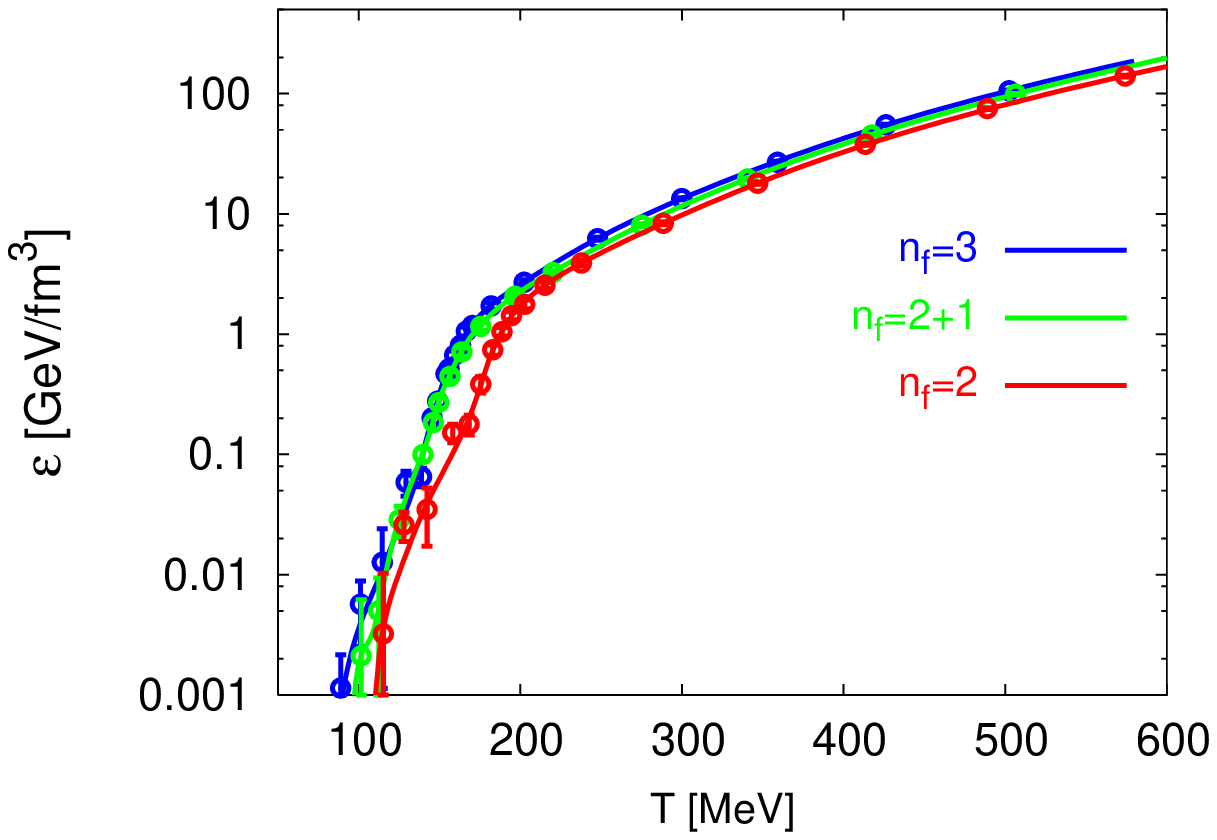}
\includegraphics{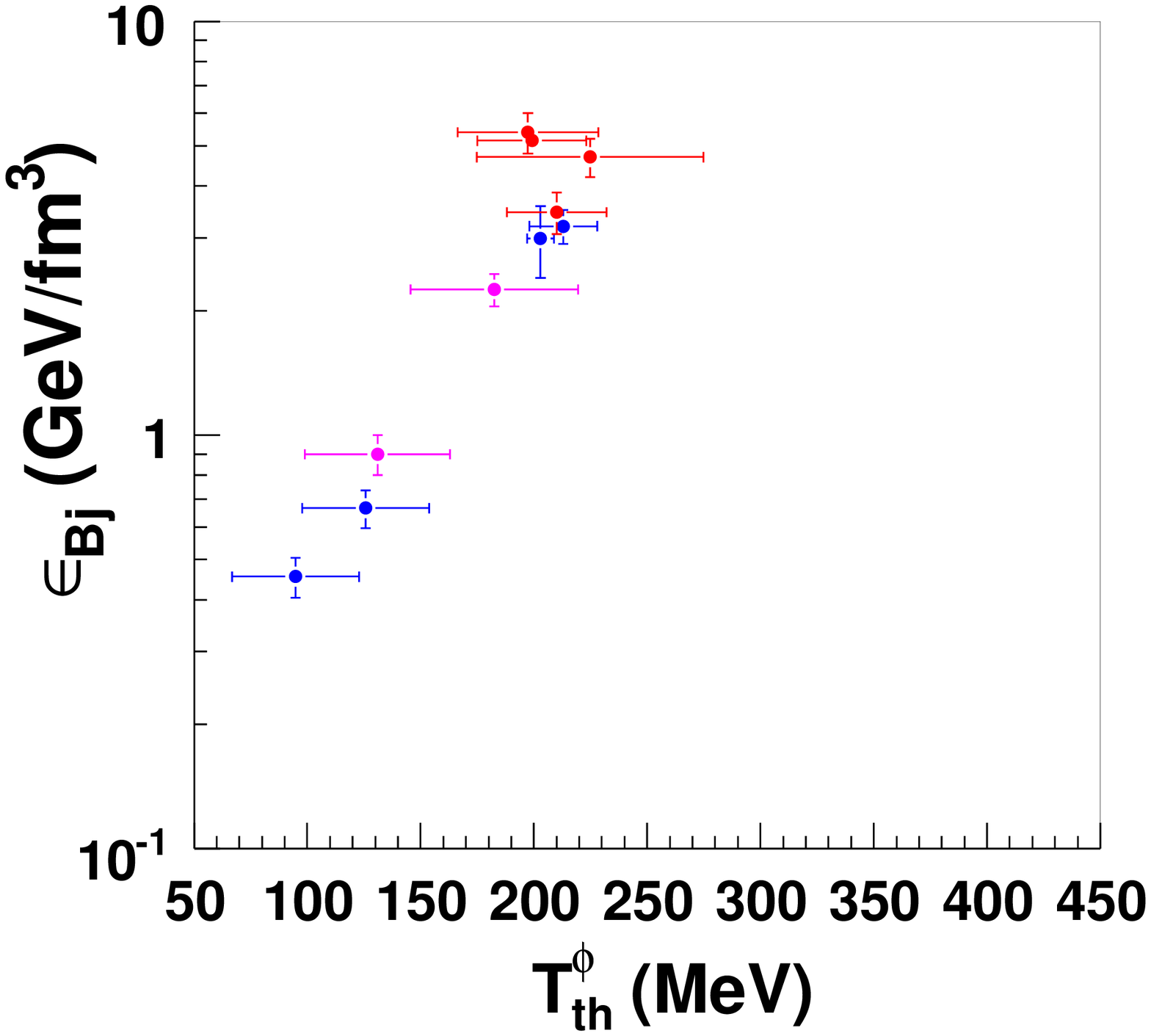}
\includegraphics{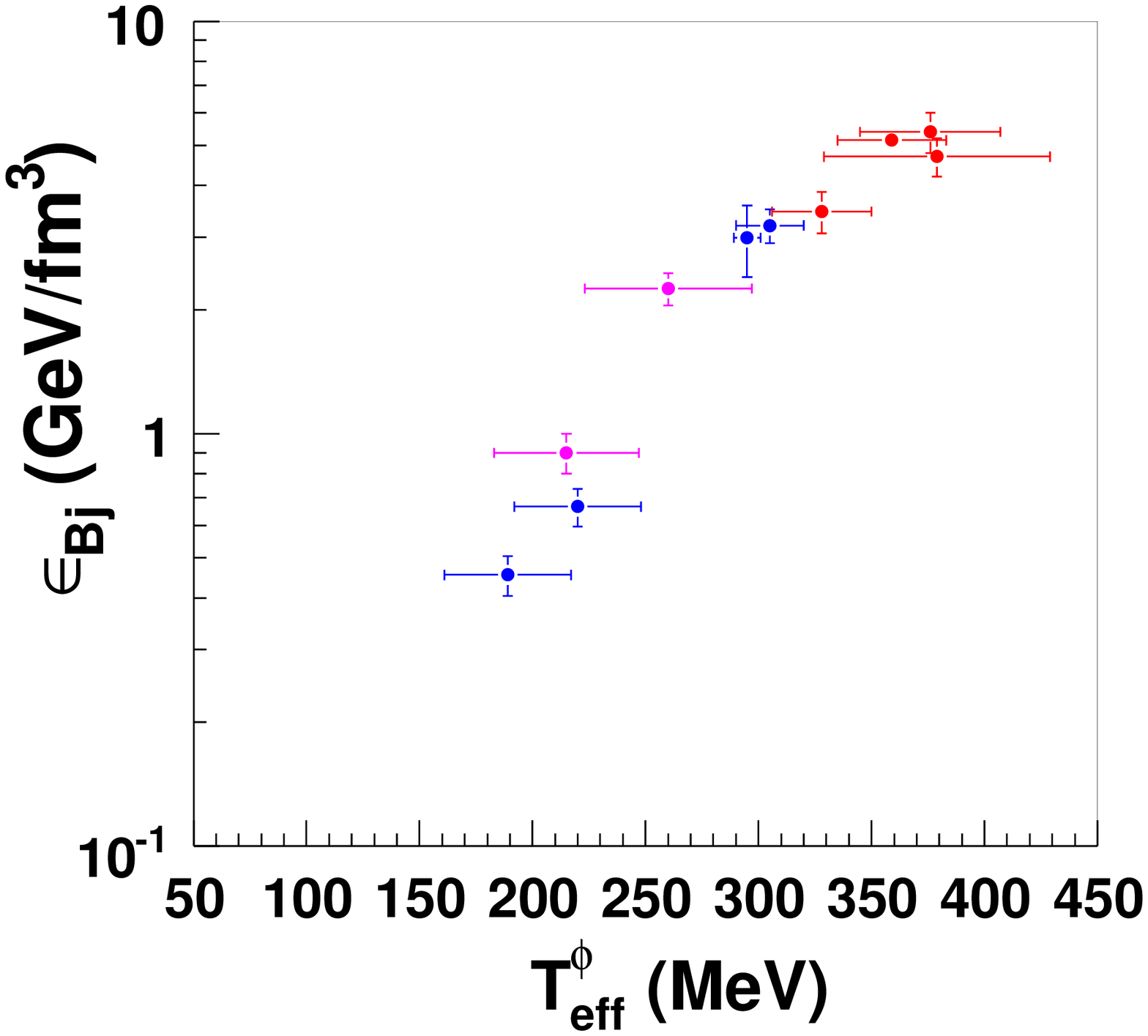}}
\caption{Energy density as a function of temperature: lattice plot from Ref. 
\cite{lQCD} (left).Middle and right: Energy density as a function of $T_{th}$ (middle)
 and $T_{eff}$ (right) for the $\phi$.}
\label{withLattice}
\end{center}
\end{figure}

\noindent
~~~In the following we  compare experimental data with  lattice QCD predictions.
In figure \ref{withLattice} 
the lattice prediction for the energy density is shown as a function
of the temperature (left). This is compared to $\epsilon_{Bj}$ 
as a function of $T^{\phi}_{th}$ (middle) and $T^\phi_{eff}$  (right) extracted  from the 
slope of $\phi$ spectra.
First we discuss the middle plot.
The lattice prediction is for zero baryochemical potential
therefore correspondig to the cross over region.
The figure in the middle (data) is at non-zero $\mu_B$ and shows the variation of the initial  energy
density with $T^\phi_{th}$
temperature, which is measured at a later time than the energy density, namely at the thermal freeze out of
the $\phi$.
That is,
 the two variables in the data are a measure of the system at different times, while the lattice
estimate is independent of time.
This temperature
is expected to be always below $T_c$ and does not reflect the initial $T$, 
which may exceed $T_c$ depending on the collision energy.
This
 plot can be directly compared to figure 
11 of \cite{strangeborder}, where the energy density from data
 has been studied as a function of T,
while both x and y axis were estimated at the same time, namely at the hadronic chemical
freeze out time and at $\mu_{B} =0$.
It is seen that the energy density increases approaching the $T_c$ from below, from
which both the $T_c$ and critical exponents can be extracted \cite{strangeborder}.

\noindent
~~~If the transition occurs at $T_c \sim 200$ MeV as expected,
the temperature of $\phi$ at the thermal freeze out
should saturate below and near $T_c$, for
all values of the initial 
density (up to infinity), as seen in figure \ref{withLattice} (middle panel).
We observe a saturation in the value of $T^\phi_{th}$ near $T_C$, because 
 $T^{\phi}_{th}\leq T_c$  as mentioned earlier. To measure temperatures above $T_c$ photons 
and dileptons are useful probes.  

\noindent
To avoid this saturation and to study an equivalent of the
initial temperature, we  use in the following figure (right) the
inverse slope of $\phi$ instead of its temperature at hadronic freeze out.
Doing
 that, we take advantage of the possible relationship between collective
transverse flow and initial pressure.

\noindent
~~~We now discuss figure \ref{withLattice}
(right panel), where the initial Bjorken energy
density is shown 
as a function of the effective temperature  of the  $\phi$
for collisions from AGS, SPS to RHIC.
Now both the x and y axis are reflecting parameters at initial times
therefore this plot is more appropriate to be compared to the lattice results.
The
 effective temperature  of the  $\phi$
here is a sum of a thermal freeze out temperature, which is expected
to be always below $T_c$,  and
a non-thermal component
due to transverse flow, which relates to the initial pressure
and reflects the initial conditions above $T_c$.
Therefore the variables in the two plots compared here are not exactly the same
but they are correlated.
Further 
analysis is needed to compare the data  to lattice e.g. using exactly the same
variables in both estimates.
A study can be also done involving the elliptic flow $v_2$ and the transverse
flow velocity  $\beta_T$ as a function of the initial energy density.
Also a hydrodynamic calculation of the discussed variables is of interest.
The above items are work in progress.

\section{Summary}
In summary,  a first analysis of experimental data on the inverse slope parameter of $\phi$ mesons 
as a function of $\sqrt{s_{NN}}$ \& energy density 
from AGS, SPS to RHIC,
and their comparison to simple thermodynamic expectations, suggest a first order phase transition
as predicted by lattice QCD at non-zero baryochemical potentials.
The
 mixed phase of quarks, gluons and hadrons, is found to stretch between $\epsilon_{Bj}$ 
of $\sim$ 1 and 3.2 GeV/$fm^3$. The latent heat density in a first order phase transition is  
 $\epsilon_Q(T_c)-\epsilon_H(T_c)=4B$,
where $\epsilon_Q (\epsilon_H)$ is the energy density of QGP (hadrons)
at $T_c$ and $B$ is bag constant. With $\epsilon_Q(T_c)\sim 3.2$ GeV/fm$^3$ and 
$\epsilon_H(T_c)\sim 1$ GeV/fm$^3$,
 we get a reasonable value for $B^{1/4}\sim 250$ MeV. If we ignore $B$, then
$\epsilon_Q/\epsilon_H \sim g_Q/g_H\sim 3.2$ also a reasonable number comparable
to results from lattice QCD \cite{karsch}, here $g_Q (g_H)$ is statistical degeneracy of QGP (hadrons).
The 
plateau and subsequent increase of the inverse slope parameter of $\phi$
 above $\epsilon_{Bj}$  3.2 GeV/$fm^3$
is in agreement with data from pions, kaons and protons, and is
  as well observed in the transverse flow velocity $\beta_T$
reflecting the behaviour of the initial pressure.
A first attempt to compare data to lattice QCD predictions is made. 

\noindent
~~~The
 above results, while supporting the order of the transition predicted by lattice 
at large $\mu_B$
up to $\epsilon_{Bj}\sim 3.2$  GeV/fm$^3$,
they  do not exclude a change of order of the transition
at smaller baryochemical pontials.
Which points to further work, towards  mapping out the order of the QCD phase transition
as a function of energy and baryochemical potential.

\ack
We thank Drs. M. Floris, V. Friese, D. Jouan, A. De Falco for providing SPS data.
One of us (RS) would like to thank Dr. Y.P. Viyogi for stimulating discussions
at the beginning of this work.


\section*{References}
\begin{thebibliography}{90}
\bibitem{lQCD} Miller D E 2007 \pr{\bf 443} 55
\bibitem{vanHove} Van Hove L 1982 \plb{\bf 118} 138
\bibitem{bm} Mohanty B {\it et al.} 2003 \prc{\bf 68} 021901(R)
\bibitem{marek} Gorenstein M I, Gazdzicki M and Bugaev K A  2003 \plb{\bf 567} 175
\bibitem{kodama} Y Hama et al, Acta Phys. Pol. B, 35, (2004).
\bibitem{bedanga_sqm2008} B. Mohanty, this conference proceedings.
\bibitem{bjorken} Bjorken J D 1983 \prd{\bf 27} 140
\bibitem{agsPhi} Akiba Y {\it et al.} (E-802 Collaboration) 1996 \prl{\bf 76} 2021,\\
Back B B {\it et al.} (E917 Collaboration) 2004 \prc{\bf 69} 054901
\bibitem{floris} Floris M {\it et al.} (NA60 Collaboration) 2007 \epjc{\bf 49} 255
\bibitem{na38} Abreu M C {\it et al.} (NA38 Collaboration) 1996 \plb{\bf 368} 239
\bibitem{na50} Alessandro B {\it et al.} (NA50 Collaboration) 2003 \plb{\bf 555} 147
\bibitem{na49} Afanasiev S V {\it et al.} (NA50 Collaboration) 2000 \plb{\bf 491} 59
\bibitem{spsPhi} Abreu M C {\it et al.} (NA38 Collaboration) 2005 \epjc{\bf 44} 375, Alt C {\it et al.} (NA49 Collaboration) 2005 \prl{\bf 94} 052301,2008 \prl{\bf 78} 044907
\bibitem{rhicPhi} Abelev B I {\it et al.} (STAR Collaboration) 2008
{\it Preprint} 0809.4737, \\Adler S S {\it et al.} (PHENIX Collaboration)
2005 \prc{\bf 72} 014903
\bibitem{teffRef} Heinz U 2001 \npa{\bf 685} 414, Schnedermann E, Sollfrank J and Heinz U 1993 \prc{\bf 48} 2462
\bibitem{nu} Xu N 2005 \npa {\bf 751} 109c
\bibitem{starPRL} Adams J {\it et al.} (STAR Collaboration) 2004 \prl{\bf 92} 112301
\bibitem{lokesh_sqm2008} L. Kumar, STAR coll., this conference proceedings.
\bibitem{strangeborder} S. Kabana, Eur. Phys. J. C21:545-555, 2001, hep-ph/0104001,
 S. Kabana, P. Minkowski, New J. Phys. 3:4, 2001, hep-ph/0010247.
\bibitem{karsch} Karsch F. 2002 \npa{\bf 698} 199c
\end{thebibliography}
\end{document}